\documentclass[referee,epjc3]{svjour3}          

\RequirePackage[T1]{fontenc}

\smartqed  

\RequirePackage{graphicx}
\RequirePackage{flushend}
\RequirePackage[numbers,sort&compress]{natbib}
\RequirePackage[colorlinks,citecolor=blue,urlcolor=blue,linkcolor=blue]{hyperref}
\RequirePackage{epstopdf}
\RequirePackage{slashed}
\usepackage{amsmath}
\usepackage{times}
\RequirePackage{bm}
\RequirePackage{tabularx}

\journalname{Eur. Phys. J. C.}

\begin{document}

\title{Baryon chiral perturbation theory with Wilson fermions up to $\mathcal{O}(a^2)$ and discretization effects of latest  $n_f=2+1$ LQCD octet baryon masses}

\author{Xiu-Lei Ren\thanksref{addr1,addr2}
        \and
        Li-Sheng Geng\thanksref{e1,addr1,addr2,addr3}
        \and
        Jie Meng\thanksref{addr1,addr2,addr4,addr5} 
}
%
\thankstext{e1}{e-mail: lisheng.geng@buaa.edu.cn}

\institute{School of Physics and Nuclear Energy Engineering, Beihang University, Beijing 100191, China\label{addr1}
           \and
           International Research Center for Nuclei and Particles in the Cosmos, Beihang University, Beijing 100191, China\label{addr2}
           \and
           Physik Department, Technische Universit\"{a}t M\"{u}nchen, Garching D-85747, Germany\label{addr3}
           \and
           State Key Laboratory of Nuclear Physics and Technology, School of Physics, Peking University, Beijing 100871, China\label{addr4}
           \and
           Department of Physics, University of Stellenbosch, Stellenbosch 7602, South Africa\label{addr5}
}

\date{Received: date / Accepted: date}

\titlerunning{Baryon chiral perturbation theory with Wilson fermions up to $\mathcal{O}(a^2)$ and  discretization effects of latest $n_f=2+1$ LQCD octet baryon masses}

\maketitle


\begin{abstract}
We construct the chiral Lagrangians relevant in studies of the ground-state octet baryon masses up to $\mathcal{O}(a^2)$ by taking into account discretization effects and calculate the masses up to $\mathcal{O}(p^4)$  in the extended-on-mass-shell scheme. As an application, we study the latest $n_f=2+1$ LQCD data on the ground-state octet baryon masses
from the PACS-CS, QCDSF-UKQCD, HSC, and NPLQCD Collaborations. It is shown that
the discretization effects for the studied LQCD simulations are at the order of  one to two percent for lattice spacings up to $0.15$ fm and the pion mass up to 500 MeV.
\end{abstract}

\section{Introduction}\label{SecI}

Over the past decade, lattice quantum chromodynamics (LQCD) has become an indispensable tool in studies of the non-perturbative regime of  QCD from first principles~\cite{Wilson:1974sk,Gattringer2010}.  As a numerical solution of QCD in the discrete Euclidean space-time in a finite hypercube, its main input parameters are the quark masses $m_q$, the lattice box size $L$, and the lattice spacing $a$. Because computing time increases dramatically with decreasing quark masses, most past simulations have been performed with larger-than-physical light-quark masses. As a result, LQCD simulations  require multiple extrapolations to the continuum ($a\rightarrow 0$), to infinite space-time ($L\rightarrow \infty$), and to the physical  point with physical quark masses ($m_{q}\rightarrow m_{q}^{\rm phys.}$). For many observables, these extrapolations have led to uncertainties comparable to or even larger than the inherent statistical uncertainties. Recently,   simulations with physical light-quark masses have become available (see, e.g., Refs.~\cite{Durr:2010aw,Bazavov:2012xda}), which (will) largely reduce the systematic uncertainties related to chiral extrapolations to the physical light-quark masses.

Chiral perturbation theory (ChPT), as a low-energy effective field theory of QCD, provides another indispensable tool to understand QCD in the
non-perturbative regime~\cite{Weinberg:1978kz,Gasser:1983yg,Gasser:1984gg,Gasser:1987rb,Leutwyler:1994fi, Bernard:1995dp,Pich:1995bw,Ecker:1994gg,Pich:1998xt,Bernard:2006gx,Bernard:2007zu,Scherer2012b}.
It has long been employed to perform chiral extrapolations of and to study finite-volume corrections to LQCD simulations. Both of them are important for LQCD simulations. On the other hand, LQCD simulations with varying light-quark masses and lattice volume are extremely useful to help to fix the (sometimes many) unknown low-energy constants (LECs) of ChPT, which otherwise are difficult if not impossible to be determined. To apply ChPT to the study of LQCD simulations, in principle, one should first take the continuum limit of  LQCD data, since ChPT describes the continuum QCD and is not valid for nonzero lattice spacing. However, nowadays it is a common practice to assume that lattice spacing artifacts for current LQCD setups of $a\approx 0.1$ fm are small and can be treated as systematic uncertainties.

In order to study discretization effects on LQCD simulations, one can first write down Symanzik's effective field theory~\cite{Symanzik:1983dc,Symanzik:1983gh,Sheikholeslami:1985ij,Luscher:1996sc}, a continuum effective  field theory (EFT) which describes the lattice field theory close to the continuum limit, and then one can extend ChPT to be consistent with this EFT with additional symmetry breaking parameters. In this way, the chiral expansion results can naturally encode lattice spacing effects (see, e.g. Ref.~\cite{Bar:2004xp}). Sharpe and Singleton~\cite{Sharpe:1998xm} and Lee and Sharpe~\cite{Lee:1999zxa} first extended ChPT to include finite lattice spacing effects up to $\mathcal{O}(a)$ for Wilson fermions~\cite{Wilson:1974sk} (WChPT) and staggered fermions~\cite{Kogut:1974ag,Susskind:1976jm} (SChPT), respectively.  Later, Munster and Schmidt~\cite{Munster:2003ba} applied  WChPT to the study of discretization artifacts of twisted mass fermions (tmChPT)~\cite{Frezzotti:1999vv,Frezzotti:2000nk}.

In the past decade, discretization effects on the ground-state meson/baryon properties, such as masses, decay constants, electromagnetic form factors, etc., have been extensively studied in WChPT.\footnote{We focus in this work on  WChPT, but it should be noted that similar studies have been performed in  SChPT~\cite{Aubin:2003mg,Aubin:2003uc,Sharpe:2004is,
Tiburzi:2005is,Aubin:2004xd,Aubin:2005aq,Aubin:2007mc,Bernard:2013qwa} and tmChPT~\cite{Scorzato:2004da,Sharpe:2004ps,Sharpe:2004ny,WalkerLoud:2005bt,
Sharpe:2005rq,Buchoff:2008hh,Munster:2011gh}.} In the mesonic sector, the masses and decay constants of the Nambu-Glodstone mesons were first studied up to  $\mathcal{O}(m_q^2)$ and $\mathcal{O}(a)$  for the Wilson action~\cite{Rupak:2002sm} and for the mixed action~\cite{Bar:2002nr}, where Wilson sea quarks and Ginsparg-Wilson valence quarks are employed.
These studies were subsequently extended to  next-to-leading order (up to $\mathcal{O}(a^2)$)~\cite{Aoki:2003yv,Bar:2003mh}. In the one-baryon sector, a systematic study of the nucleon properties up to $\mathcal{O}(a)$ was first performed by Beane and Savage for both the mixed and the unmixed action~\cite{Beane:2003xv}. The electromagnetic properties of the octet mesons as well as of the octet and decuplet baryons were also studied up to $\mathcal{O}(a)$ for both the mixed and the unmixed action~\cite{Arndt:2004we}. Discretization effects on the nucleon and $\Delta$ masses~\cite{Tiburzi:2005vy} as well as on the vector meson masses~\cite{Grigoryan:2005zj} were also studied up to $\mathcal{O}(a^2)$.
The EFT for the anisotropic Wilson lattice action has been formulated up to $\mathcal{O}(a^2)$~\cite{Bedaque:2007xg} as well. In this context, it is interesting to note that recently several attempts have been made to determine the unknown LECs of  WChPT~\cite{Damgaard:2010cz,Akemann:2010em,Hansen:2011kk,Hansen:2011mc,Herdoiza:2013sla}.

In the past few years, fully dynamical $n_f=2+1$ simulations in the one-baryon sector have become available. The ground-state octet baryon masses might be one of the simplest observables to simulate in such a setting and serve as a benchmark for more sophisticated studies~\cite{Durr:2008zz,Aoki:2008sm,WalkerLoud:2008bp, Lin:2008pr,Alexandrou:2009qu,Aoki:2009ix,Bietenholz:2010jr,Bietenholz:2011qq,Beane:2011pc}. Many theoretical studies have been performed not only to understand the chiral extrapolations of and the finite-volume corrections to these simulations,  but also to determine the many unknown LECs appearing in ChPT up to next-to-next-to-next-to-leading order (N$^3$LO)~\cite{WalkerLoud:2008bp,Ishikawa:2009vc,Young:2009zb,MartinCamalich:2010fp,Geng:2011wq,Semke:2011ez, Semke:2012gs,Lutz:2012mq,Bruns:2012eh,Ren:2012aj,Ren:2013dzt}. In Ref.~\cite{Ren:2012aj}, it is shown that the covariant baryon chiral perturbation theory (BChPT) together with the extended-on-mass-shell (EOMS) scheme~\cite{Gegelia:1999gf,Fuchs:2003qc} can describe reasonably well all the $n_f=2+1$ LQCD data. Nevertheless, discretization effects
are ignored in all these studies, with the argument that they should be small.\footnote{In Ref.~\cite{Alvarez-Ruso:2013fza}, Alvarez-Ruso {\it et al.} performed a phenomenological study of the continuum extrapolation of the LQCD simulations of the nucleon mass by considering only
$\mathcal{O}(a^2)$ terms, and they showed that  finite-volume corrections and finite lattice spacing effects are of similar size. In
our present work we will see that they are indeed of similar size, but the $\mathcal{O}(a m_q)$ contributions are larger than the $\mathcal{O}(a^2)$ ones.}

In this work, we aim to study the discretization effects of the  LQCD simulations of the ground-state octet baryon masses up to $\mathcal{O}(a^2)$ in covariant BChPT with the EOMS renormalization scheme. Although most of the LQCD simulations are performed at a single lattice spacing, a combination of the results from different collaborations enables one to examine finite lattice spacing effects by performing a global study. We limit ourselves to the unmixed action and, therefore, we will study those simulations based on the $\mathcal{O}(a)$-improved Wilson action~\cite{Sheikholeslami:1985ij}, i.e., those of the PACS-CS~\cite{Aoki:2008sm}, QCDSF-UKQCD~\cite{Bietenholz:2011qq}, HSC~\cite{Lin:2008pr}, and NPLQCD~\cite{Beane:2011pc} Collaborations.

The paper is organized as follows. In Sect.~\ref{SecII}, the Symanzik action up to $\mathcal{O}(a^2)$ is briefly introduced and the $a$-dependent chiral Lagrangians relevant to the study of the ground-state octet baryon masses  are constructed. In Sect.~\ref{SecIII}, the discretization effects on the ground-state octet baryon masses are formulated up to $\mathcal{O}(a^2)$ for Wilson fermions. As an application, we then perform  a simultaneous fit of the LQCD octet baryon masses and study the discretization effects. A short summary is given in Sect.~\ref{SecIV}.

\section{BChPT at finite lattice spacing}\label{SecII}
In this section, we briefly review the continuum effective action up to and including $\mathcal{O}(a^2)$. We will follow closely the procedure and notations of Ref.~\cite{Tiburzi:2005vy} and construct for the first time the chiral Lagrangians incorporating a finite lattice spacing for the Wilson action in the $u$, $d$, and $s$ three-flavor one-baryon sector.

\subsection{Continuum effective action}

Close to the continuum limit,  LQCD can be described by an effective action,  the `Symanzik action'~\cite{Symanzik:1983dc,Symanzik:1983gh}, which is expanded in powers of the lattice spacing $a$ as
\begin{eqnarray}
  S_{\rm eff} &=& S_0 + a S_1 + a^2 S_2 + \cdots\nonumber\\
              &=& \int d^4x (\mathcal{L}^{(4)} + a\mathcal{L}^{(5)} + a^2\mathcal{L}^{(6)} + \cdots),
\end{eqnarray}
where $\mathcal{L}^{(4)}$ is the normal (continuum) QCD Lagrangian and the two new terms $\mathcal{L}^{(5)}$ and $\mathcal{L}^{(6)}$ are introduced to include the discretization effects of LQCD. The Lagrangian $\mathcal{L}^{(5)}$ contains chiral breaking terms only, while $\mathcal{L}^{(6)}$ contains both chiral invariant and breaking terms.
In the $u$, $d$, and $s$ three-flavor sector, the QCD Lagrangian is
\begin{equation}
  \mathcal{L}^{(4)} = \bar{\psi}(i\slashed D-\mathcal{M})\psi,
\end{equation}
where the quark masses are encoded in a diagonal matrix $\mathcal{M}={\rm diag}(m_l, m_l, m_s)$ in the isospin limit ($m_u=m_d\equiv m_l$),
and $\slashed{D}=D_\mu \gamma^\mu$ with $D_\mu$ the covariant derivative.

At $\mathcal{O}(a)$, there is only the Pauli term left by using the equations of motion to redefine the effective fields~\cite{Luscher:1996sc}
\begin{equation}
  a\mathcal{L}^{(5)} = ac_{\rm SW} \bar{\psi}\sigma^{\mu\nu}G_{\mu\nu}\omega_q\psi,
\end{equation}
where $ G_{\mu\nu}=[D_\mu,D_\nu]$ and $c_{\rm SW}$ is the Sheikholeslami-Wohlert (SW)~\cite{Sheikholeslami:1985ij} coefficient that must be determined numerically. The $\omega_q~(q=u,~d,~s)$ is a constant which is determined by the kind of lattice fermions employed in LQCD simulations: $\omega_q=1$ for Wilson fermions~\cite{Wilson:1974sk} and $\omega_q=0$ for Ginsparg-Wilson (GW) fermions~\cite{Ginsparg:1981bj}. Similar to the quark masses, the $\omega_q$'s are usually collected in the Wilson matrix $\mathcal{W}={\rm diag}(\omega_l, \omega_l, \omega_s)$ with conserved isospin symmetry ($\omega_u=\omega_d\equiv \omega_l$). This term breaks chiral symmetry in precisely the same way as the quark mass term. It should be noted that the Pauli term can be canceled by adding the clover term to the lattice action~\cite{Bar:2003mh}, resulting in the $\mathcal{O}(a)$-improved Wilson fermion action~\cite{Sheikholeslami:1985ij,Luscher:1996sc,Luscher:1996ug,Frezzotti:2003ni}.

Up to $\mathcal{O}(a^2)$, the Symanzik action for Wilson fermions has been extensively studied in Refs.~\cite{Sheikholeslami:1985ij,Aoki:2003yv,Bar:2003mh}. In total, there are $18$ operators appearing in $\mathcal{L}^{(6)}$. They can be classified into  operators of the following five types according to whether or not they break chiral symmetry and  the $O(4)$ rotation symmetry~\cite{Tiburzi:2005vy}:
\begin{itemize}
  \item $\mathcal{L}_1^{(6)}$: quark bilinear operators that conserve chiral symmetry,
  \begin{equation}
     \bar{\psi}\slashed{D}^3\psi,\quad \bar{\psi}(D_{\mu}D_{\mu}\slashed{D} + \slashed{D}D_{\mu}D_{\mu})\psi, \quad
     \bar{\psi}D_{\mu}\slashed{D} D_{\mu}\psi.
  \end{equation}
  \item $\mathcal{L}_2^{(6)}$: quark bilinear operators that break chiral symmetry,
  \begin{equation}
    \bar{\psi}m_qD_{\mu}D_{\mu}\psi,\quad \langle m_q\rangle\bar{\psi}D_{\mu}D_{\mu}\psi,\quad
    \bar{\psi}m_qi\sigma_{\mu\nu}G_{\mu\nu}\psi,\quad \langle m_q\rangle\bar{\psi}i\sigma_{\mu\nu}G_{\mu\nu}\psi.
  \end{equation}
  \item $\mathcal{L}_3^{(6)}$: four-quark operators that conserve chiral symmetry,
    \begin{equation}
    (\bar{\psi}\gamma_{\mu}\psi)^2,\quad (\bar{\psi}\gamma_{\mu}\gamma_5\psi)^2,\quad (\bar{\psi}t^a\gamma_{\mu}\psi)^2,\quad (\bar{\psi}t^a\gamma_{\mu}\gamma_5\psi)^2,
  \end{equation}
where $t^a$ are the SU(3) generators, $a=1,\cdots,8$.

  \item $\mathcal{L}_4^{(6)}$: four-quark operators that break chiral symmetry,
  \begin{equation}
    (\bar{\psi}\psi)^2,\quad (\bar{\psi}\gamma_5\psi)^2,\quad (\bar{\psi}\sigma_{\mu\nu}\psi)^2,\quad (\bar{\psi}t^a\psi)^2,\quad (\bar{\psi}t^a\gamma_5\psi)^2,\quad (\bar{\psi}t^a\sigma_{\mu\nu}\psi)^2.
  \end{equation}
  \item $\mathcal{L}_5^{(6)}$: quark bilinear operators that break the $O(4)$ rotation symmetry,
  \begin{equation}
    \bar{\psi}\gamma_{\mu}D_{\mu}D_{\mu}D_{\mu}\psi.
  \end{equation}
\end{itemize}

It should be noted that fermionic operators that conserve chiral symmetry first appear  at $\mathcal{O}(a^2)$.

\subsection{Wilson chiral Lagrangians }

In order to construct the chiral  Lagrangians of  WChPT, one has to write down the most general Lagrangians that are invariant under the symmetries of the continuum EFT. This can be done by following the standard procedure of spurion analysis~\cite{Aoki:2003yv,Bar:2003mh}. In practice,  in order to obtain the corresponding $a$-dependent chiral Lagrangians, one only needs to know which symmetries are broken and how~\cite{Tiburzi:2005vy}. Before writing down the chiral Lagrangians up to $\mathcal{O}(a^2)$, one has to first specify a chiral power-counting scheme, which should be enlarged to include the lattice spacing $a$.  In LQCD simulations, the following hierarchy of energy scales is satisfied:
\begin{equation}
  m_q \ll \Lambda_{\rm QCD} \ll \frac{1}{a}.
\end{equation}
If one assumes that the size of the chiral symmetry breaking due to the light-quark masses and the discretization effects are of comparable size, as done in Refs.~\cite{Beane:2003xv,Bar:2003mh,Tiburzi:2005vy}, one has the following expansion parameters:
\begin{equation}
  p^2\sim \frac{m_q}{\Lambda_{\rm QCD}} \sim a\Lambda_{\rm QCD},
\end{equation}
where $p$ denotes a generic small quantity and $\Lambda_{\rm QCD}\approx 300$ MeV denotes the typical low energy scale of QCD. Up to $\mathcal{O}(a^2)$, the $a$-dependent chiral Lagrangians contain terms of $\mathcal{O}$ $(a$, $am_q$, $a^2)$ and can be written as
\begin{equation}
  \mathcal{L}_{a}^{\rm eff} = \mathcal{L}_a^{(1)} + \mathcal{L}_a^{(2)},
\end{equation}
where
\begin{eqnarray}
  \mathcal{L}_a^{(1)} &=& \mathcal{L}^{\mathcal{O}(a)} + \mathcal{L}^{\mathcal{O}(am_q)},\\
  \mathcal{L}_a^{(2)} &=& \mathcal{L}_1^{\mathcal{O}(a^2)} + \mathcal{L}_2^{\mathcal{O}(a^2)} + \mathcal{L}_3^{\mathcal{O}(a^2)} + \mathcal{L}_4^{\mathcal{O}(a^2)}+\mathcal{L}_5^{\mathcal{O}(a^2)},
\end{eqnarray}
and $\mathcal{L}_{i}^{\mathcal{O}(a^2)} (i=1,\ldots, 5)$ are the five classes of chiral Lagrangians corresponding to the previous five types of operators appearing in the Symanzik action at $\mathcal{O}(a^2)$.

\begin{figure}[t]
  \centering
  \includegraphics[width=\textwidth]{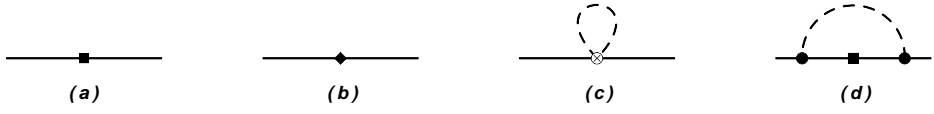}\\
  \caption{Feynman diagrams contributing to the $a$-dependence of octet baryon masses up to $\mathcal{O}(a^2)$. The solid lines represent octet baryons and the dashed lines denote pseudoscalar mesons. The boxes (diamonds) indicate the $\mathcal{O}(a)$ ($\mathcal{O}(a^2)$) vertices. The circle-cross is an insertion from the $\mathcal{L}^{\mathcal{O}(a)}$. The wave function renormalization diagrams are not explicitly shown but included in the calculation.}\label{Fig:1}
\end{figure}

The chiral Lagrangian at $\mathcal{O}(a)$ can be written as
\begin{equation}
  \mathcal{L}^{\mathcal{O}(a)}=\bar{b}_0\langle\bar{B}B\rangle\langle\rho_+\rangle + \bar{b}_D\langle\bar{B}[\rho_+, B]_-\rangle + \bar{b}_F\langle\bar{B}[\rho_+, B]_+\rangle,
\end{equation}
where $\bar{b}_0$, $\bar{b}_D$, and $\bar{b}_F$ are the unknown LECs of dimension mass$^{-1}$, $\langle X\rangle$ stands for the trace in flavor space, $\rho_+=u^{\dag}\rho u^{\dag}+u\rho^{\dag}u$ with $u=\sqrt{U}={\rm exp}(i\phi/(2F_{\phi}))$,\footnote{The operator $\rho_+$ transforms under chiral rotation (R), parity transformation (P), charge conjugation transformation (C) and hermitic conjugation transformation in the following way: $\rho_+ \xrightarrow{R} h\rho_+ h^\dag$ with $h\in\mathrm{SU(3)}_V$, $\rho_+\xrightarrow{P} \rho_+$, $\rho_+\xrightarrow{C} \rho_+^T$, and $\rho_+\xrightarrow{h.c.} \rho_+$.}
$\phi$ and $B$ are the usual SU(3) matrix representation of the pseudoscalar mesons and of the octet baryons, respectively. The coefficient $F_{\phi}$ is the pseudoscalar decay constant in the chiral limit. The matrix $\rho$ is related to the Wilson matrix via~\cite{Rupak:2002sm}
\begin{equation}
\rho = 2ac_{\rm SW}W_0\mathcal{W},
\end{equation}
which introduces explicit chiral symmetry breaking because of the finite lattice spacing~$a$. The constant $W_0=-\langle0|\bar{q}\sigma_{\mu\nu}G_{\mu\nu}q|0\rangle/F_{\phi}^2$ is an unknown dimensional quantity that is related to the scale $\Lambda_{\chi}$.

The  $\mathcal{O}(am_q)$ Lagrangian has the following form:
\begin{eqnarray}\label{Lamq}
  \mathcal{L}^{\mathcal{O}(am_q)}&=& \bar{b}_1\langle\bar{B}\chi_+\rho_+ B\rangle + \bar{b}_2\langle\bar{B}\chi_+ B\rho_+\rangle + \bar{b}_3\langle\bar{B}\rho_+ B\chi_+\rangle + \bar{b}_4\langle\bar{B}B\chi_+\rho_+\rangle\nonumber\\
  && + \bar{b}_5\langle\bar{B}\chi_+\rangle\langle\rho_+ B\rangle + \bar{b}_6\langle\bar{B}\rho_+\rangle\langle\chi_+B\rangle + \bar{b}_7\langle\bar{B}[\chi_+,B]\rangle\langle\rho_+\rangle \nonumber\\
  &&+\bar{b}_8\langle\bar{B}\{\chi_+,B\}\rangle\langle\rho_+\rangle+ \bar{b}_9\langle\bar{B}[\rho_+,B]\rangle\langle\chi_+\rangle +\bar{b}_{10}\langle\bar{B}\{\rho_+,B\}\rangle\langle\chi_+\rangle\nonumber\\
  &&   + \bar{b}_{11}\langle\bar{B}B\rangle\langle\chi_+\rangle\langle\rho_+\rangle + \bar{b}_{12}\langle\bar{B}B\rangle\langle\chi_+\rho_+\rangle,
\end{eqnarray}
where $\bar{b}_{1,\ldots,12}$ are unknown LECs of dimension mass$^{-3}$ and $\chi_+=u^{\dag}\chi u^{\dag} + u\chi^{\dag} u$, where
$\chi = 2B_0 \mathcal{M}$
accounts for explicit chiral symmetry breaking with $B_0=-\langle0|\bar{q}q|0\rangle/F_{\phi}^2$.
One can eliminate the $\bar{b}_3$ term by use of  the following identity valid for any $3\times3$ matrix $A$ derived from the Cayley-Hamilton identity~\cite{Zhang:2007qp}:
\begin{eqnarray}
  &&\sum\limits_{\rm perm=6}\langle A_1A_2A_3A_4\rangle-\sum\limits_{\rm perm=8}\langle A_1A_2A_3\rangle\langle A_4\rangle - \sum\limits_{\rm perm=3}\langle A_1A_2\rangle \langle A_3A_4\rangle\nonumber\\ &&+\sum\limits_{\rm perm=6}\langle A_1A_2\rangle\langle A_3\rangle\langle A_4\rangle
  -\langle A_1\rangle\langle A_2\rangle\langle A_3\rangle\langle A_4\rangle =0,
\end{eqnarray}
where `perm' stands for permutation number. In the end, there are $11$ independent terms left.

At $\mathcal{O}(a^2)$, the previous five operators in the Symanzik action can be mapped into the EFT with five classes of chiral Lagrangians $\mathcal{L}_i^{\mathcal{O}(a^2)}~(i=1,\ldots,5)$. Following the notation of Ref.~\cite{Tiburzi:2005vy},  the first class of chiral  Lagrangians can be written as
\begin{equation}\label{Eq:a2-1}
  \mathcal{L}_1^{\mathcal{O}(a^2)} = a^2 c_{\rm SW}^2W_0^2\left[
  \bar{c}_1\langle\bar{B}B\rangle + \bar{c}_2\langle\mathcal{O}_+\rangle\langle \bar{B}B\rangle + \bar{c}_3\langle\bar{B}[\mathcal{O}_+, B]_+\rangle + \bar{c}_4\langle\bar{B}[\mathcal{O}_+,B]_-\rangle\right],
\end{equation}
where the operator $\mathcal{O}_+$ is defined as
\begin{equation}
  \mathcal{O}_+= 2\left[u^{\dag}(\mathcal{W}-\overline{\mathcal{W}})u + u(\mathcal{W}-\overline{\mathcal{W}})u^{\dag}\right],
\end{equation}
with $\overline{\mathcal{W}}=1-\mathcal{W}={\rm diag}(1-\omega_l,1-\omega_l,1-\omega_s)$, and $\bar{c}_{1,\ldots,4}$ are the unknown LECs of dimension mass$^{-3}$.

Because the second type of operators have an insertion of the quark mass $m_q$, the chiral order of the corresponding chiral Lagrangians is at least $\mathcal{O}(p^6)$, which is beyond the present work and will not be shown.

There are seven independent terms in the third class of chiral Lagrangians
\begin{eqnarray}\label{c3}
  \mathcal{L}_3^{\mathcal{O}(a^2)} &=& a^2c^2_{\rm SW}W_0^2\left[\bar{e}_1 \langle\bar{B}[\mathcal{O}_+, [\mathcal{O}_+, B]]\rangle + \bar{e}_2 \langle\bar{B}[\mathcal{O}_+, \{\mathcal{O}_+, B\}]\rangle\right.\nonumber\\
  \qquad && + \bar{e}_3 \langle\bar{B}\{\mathcal{O}_+,\{\mathcal{O}_+, B\}\}\rangle + \bar{e}_4 \langle\bar{B}\mathcal{O}_+\rangle\langle\mathcal{O}_+B\rangle\nonumber\\
  \qquad && + \bar{e}_5 \langle\bar{B}[\mathcal{O}_+, B]\rangle \langle\mathcal{O}_+\rangle
  + \bar{e}_6 \langle\bar{B}\{\mathcal{O}_+, B\}\rangle \langle\mathcal{O}_+\rangle \nonumber\\
  \qquad && \left.+\bar{e}_7 \langle\bar{B}B\rangle\langle\mathcal{O}_+\rangle^2   +\bar{e}_8 \langle\bar{B}B\rangle\langle\mathcal{O}_+^2\rangle\right],
\end{eqnarray}
where the $\bar{e}_i$ are the unknown LECs of dimension mass$^{-3}$. Furthermore, we can eliminate the $\bar{e}_6 $ term by use of the Cayley-Hamilton identity~\cite{Zhang:2007qp}:
\begin{equation}
  \langle\bar{B}\{X^2, B\}\rangle + \langle\bar{B}XBX\rangle - \frac{1}{2}\langle\bar{B}B\rangle\langle X^2\rangle - \langle\bar{B}X\rangle\langle BX\rangle=0,
\end{equation}
with $X=\mathcal{O}_+-\frac{1}{3}\langle\mathcal{O}_+\rangle$ being a $3\times 3$ traceless matrix.

Four-quark operators that break chiral symmetry can be mapped into the following chiral Lagrangian:
 \begin{eqnarray}\label{c4}
  \mathcal{L}_4^{\mathcal{O}(a^2)} &=& \bar{d}_1\langle\bar{B}[\rho_+, [\rho_+, B]]\rangle + \bar{d}_2\langle\bar{B}[\rho_+, \{\rho_+, B\}]\rangle\nonumber\\
  \qquad && + \bar{d}_3\langle\bar{B}\{\rho_+,\{\rho_+, B\}\}\rangle + \bar{d}_4\langle\bar{B}\rho_+\rangle\langle\rho_+B\rangle\nonumber\\
  \qquad && + \bar{d}_5\langle\bar{B}[\rho_+, B]\rangle \langle\rho_+\rangle
  + \bar{d}_7\langle\bar{B}B\rangle\langle\rho_+\rangle^2 \nonumber\\
  && +\bar{d}_8\langle\bar{B}B\rangle\langle\rho_+^2\rangle,
\end{eqnarray}
with the seven unknown LECs $\bar{d}_{i}$ of dimension mass$^{-3}$. Because the chiral transformation properties of $\rho_+$ and $\chi_+$ are the same,  the chiral Lagrangian has the same form as the corresponding fourth-order chiral Lagrangian of ChPT.

For the $O(4)$  breaking operators, the mapped chiral Lagrangian can be written as
\begin{eqnarray}
  \mathcal{L}_{5}^{\mathcal{O}(a^2)} &=& a^2c_{\rm SW}^2W_0^2\left[\bar{f}_1\langle\bar{B}D_{\mu}D_{\mu}D_{\mu}D_{\mu}B\rangle +\bar{f}_2\langle\mathcal{O}_+\rangle \langle\bar{B}D_{\mu}D_{\mu}D_{\mu}D_{\mu}B\rangle\right.\nonumber\\
  &&\left. +\bar{f}_3\langle\bar{B}D_{\mu}D_{\mu}D_{\mu}D_{\mu}[\mathcal{O}_+,B]_+\rangle + \bar{f}_4\langle\bar{B}D_{\mu}D_{\mu}D_{\mu}D_{\mu}[\mathcal{O}_+,B]_-\rangle \right],
\end{eqnarray}
where the $\bar{f}_i$ are the unknown LECs of dimension mass$^{-3}$. Their contributions to the octet baryon masses can be absorbed by the terms of class one, i.e., Eq.~(\ref{Eq:a2-1}).

\section{Discretization effects on the octet baryon masses}\label{SecIII}
In this section, we calculate the discretization effects on the octet baryon masses up to $\mathcal{O}(a^2)$ for the Wilson action. Then
employing the baryon masses obtained in Wilson covariant BChPT up to N$^3$LO, we estimate discretization effects of the current LQCD simulations by performing a simultaneous fit of the latest $n_f=2+1$ LQCD data, which are obtained with the $\mathcal{O}(a)$-improved Wilson action.

It should be stressed that we are not aiming at a precise determination of discretization effects on the octet baryon masses,
given the fact that most LQCD simulations are performed at a single lattice spacing. On the contrary, we would like
to get a rough estimate of discretization effects and to check whether the results of previous studies~\cite{MartinCamalich:2010fp,Geng:2011wq,Ren:2012aj,Ren:2013dzt} are robust, which have neglected these effects.
\subsection{Octet baryon masses up to $\mathcal{O}(p^4)$ }

The octet baryon masses up to N$^3$LO and with finite lattice spacing $a$ contributions up to  $\mathcal{O}(a^2)$ can be expressed as
\begin{equation}\label{Eq:mass}
  m_B = m_0 + m_B^{(2)} + m_B^{(3)} + m_B^{(4)} + m_B^{(a)},
\end{equation}
where $m_0$ is the chiral limit octet baryon mass and $m_B^{(2)}$, $m_B^{(3)}$, and $m_B^{(4)}$ correspond to the $\mathcal{O}(p^2)$, $\mathcal{O}(p^3)$, and $\mathcal{O}(p^4)$ contributions (the corresponding finite-volume corrections from loop diagrams are also included) and their explicit expressions can be found in Ref.~\cite{Ren:2012aj}. The last term $m_B^{(a)}$ denotes the discretization effects
 up to $\mathcal{O}(a^2)$. In our power-counting scheme, it contains the following three contributions:
\begin{equation}
  m_B^{(a)} = m_B^{\mathcal{O}(a)} + m_B^{\mathcal{O}(am_q)} + m_B^{\mathcal{O}(a^2)}.
\end{equation}
Here, we need to mention that virtual decuplet contributions are not explicitly included, since their effects on the chiral extrapolation and
 the finite-volume corrections are relatively small~\cite{Ren:2013dzt}.

In the case of the unmixed Wilson action, where the $u$, $d$, and $s$ quarks are all Wilson fermions, the Wilson matrix can be written as $\mathcal{W}={\rm diag}(1,1,1)$.
One can easily compute the $\mathcal{O}(a)$ contributions of the diagram ~Fig.~\ref{Fig:1}a to the octet baryon masses,
\begin{equation}\label{Eq:a1}
  m_B^{\mathcal{O}(a)} = - 4 ac_{\rm SW} W_0 (3\bar{b}_0+2\bar{b}_D),
\end{equation}
where $B=N,~\Lambda,~\Sigma,$ and $\Xi$.

\begin{table}[b]
  \centering
  \caption{Coefficients of the $\mathcal{O}(am_q)$ contributions to the octet baryon masses~(Eq.~\ref{Eq:amq}). }
  \label{Tab:amq}
  \begin{tabular}{ccc}
    \hline\hline
      & $\xi_l$ & $\xi_s$ \\
    \hline
    $N$ & $\bar{B}_1+2\bar{B}_3$ & $\bar{B}_2+\bar{B}_3$ \\
    $\Lambda$ & $\frac{1}{3}(\bar{B}_1+\bar{B}_2+6\bar{B}_3)$ & $\frac{1}{3}(2\bar{B}_1+2\bar{B}_2+3\bar{B}_3)$ \\
    $\Sigma$ & $\bar{B}_1+\bar{B}_2+2\bar{B}_3$ & $\bar{B}_3$ \\
    $\Xi$ & $\bar{B}_2+2\bar{B}_3$ & $\bar{B}_1+ \bar{B}_3$ \\
    \hline\hline
  \end{tabular}
\end{table}

The $\mathcal{O}(am_q)$ contributions can be written as
\begin{eqnarray}\label{Eq:amq}
  m_B^{\mathcal{O}(am_q)} &=& -16ac_{\rm SW}W_0B_0(\xi_l m_l + \xi_s m_s)\nonumber\\
            &=& -8ac_{\rm SW}W_0\left(\xi_lM_{\pi}^2 + \xi_s(2M_K^2-M_{\pi}^2)\right),
\end{eqnarray}
and the coefficients $\xi_l$ and $\xi_s$ are tabulated in Table~\ref{Tab:amq}. We have introduced the following combinations of LECs:  $\bar{b}_1+\bar{b}_2+3\bar{b}_7+3\bar{b}_8=\bar{B}_1$, $\bar{b}_4-3\bar{b}_7+3\bar{b}_8=\bar{B}_2$, and $2\bar{b}_{10}+3\bar{b}_{11}+\bar{b}_{12}=\bar{B}_3$. Hence, there are $3$ independent combinations.
In obtaining the above results, the light-quark masses have been replaced by the leading-order pseudoscalar meson masses: $m_l=\frac{1}{2B_0}M_{\pi}^2$ and $m_s=\frac{1}{2B_0}(2M_K^2-M_{\pi}^2)$.

The $\mathcal{O}(a^2)$ contributions are not only from the fourth-order tree-level diagram~Fig.~\ref{Fig:1}-(b), but also from the one-loop diagrams of Fig.~\ref{Fig:1}c,d
\begin{eqnarray}\label{Eq:a2}
  m_B^{\mathcal{O}(a^2)} &=& - a^2c_{\rm SW}^2W_0^2\left(\bar{C} + 16\bar{D} + 16\bar{E} \right)\nonumber\\
  && -\frac{1}{(4\pi F_{\phi})^2}ac_{\rm SW} W_0 \sum\limits_{\pi,~K,~\eta}\xi_{B,\phi}^{(c)}  H_B^{(c)}(M_{\phi})\nonumber\\
  && + \frac{1}{(4\pi F_{\phi})^2}\sum\limits_{\pi,~K,~\eta}\xi_{BB^{\prime},\phi}^{(d)}  H_{B,B^{\prime}}^{(d)}(M_{\phi}),
\end{eqnarray}
where  $\bar{C}=\bar{c}_1+4(3\bar{c}_2+2\bar{c}_3)$, $\bar{D}=4\bar{d}_3+9\bar{d}_7+3\bar{d}_8$, and $\bar{E}=4\bar{e}_3+9\bar{e}_7+3\bar{e}_8$.
We introduce $\bar{C}+16\bar{D}+16\bar{E}=16\bar{X}$ as one free LEC in the fitting process. The second line of Eq.~(\ref{Eq:a2}) is  for the contributions from the tadpole diagram of Fig.~\ref{Fig:1}c, and the corresponding coefficients $\xi_{B,\phi}^{(c)}$ are listed in Table~\ref{Tab:tada}.
The last term is for the contributions from the one-loop diagram of Fig.~\ref{Fig:1}d,
and the coefficients $\xi_{BB^{\prime},\phi}^{(d)}$ can be found in Table 5 of Ref.~\cite{Ren:2012aj}.
The loop diagrams $H_{B}^{(c)}(M_{\phi})$ and $H_{BB^{\prime}}^{(d)}(M_{\phi})$ read
\begin{eqnarray}
  H_B^{(c)}(M_{\phi}) &=& M_{\phi}^2\left[1+\ln\left(\frac{\mu^2}{M_{\phi}^2}\right)\right],\\
  H_{BB^{\prime}}^{(d)}(M_{\phi}) &=& m_B^{\mathcal{O}(a)}\left[\frac{2M_{\phi}^5}{m_0^2\sqrt{4m_0^2-M_{\phi}^2}}\arccos\left(\frac{M_{\phi}}{2m_0}\right) +\frac{M_{\phi}^4}{m_0^2}\ln\left(\frac{M_{\phi}^2}{m_0^2}\right) \right.\nonumber\\
  &&\left. +2M_{\phi}^2\ln\left(\frac{m_0^2}{\mu^2}\right) \right],
\end{eqnarray}
where the $m_B^{\mathcal{O}(a)}$ is for the leading-order discretization effects of Eq.~(\ref{Eq:a1}).

\begin{table}[h!]
  \centering
  \caption{Coefficients of the tadpole diagram contributions to the octet baryon masses~(Eq.~(\ref{Eq:a2})).}
  \label{Tab:tada}
  \begin{tabular}{ccccc}
    \hline\hline
      & $N$ & $\Lambda$ & $\Sigma$ & $\Xi$ \\
    \hline
    $\xi_{B,\pi}^{(c)}$ & $6(2\bar{b}_0+\bar{b}_D+\bar{b}_F)$ & $4(3\bar{b}_0+\bar{b}_D)$ & $12(\bar{b}_0+\bar{b}_D)$ & $6(2\bar{b}_0+\bar{b}_D-\bar{b}_F)$ \\
    $\xi_{B,K}^{(c)}$ & $4(4\bar{b}_0+3\bar{b}_D-\bar{b}_F)$ & $\frac{8}{3}(6\bar{b}_0+5\bar{b}_D)$ & $8(2\bar{b}_0+\bar{b}_D)$ & $4(4\bar{b}_0+3\bar{b}_D+\bar{b}_F)$ \\
    $\xi_{B,\eta}^{(c)}$ & $\frac{2}{3}(6\bar{b}_0+5\bar{b}_D-3\bar{b}_F)$ & $4(\bar{b}_0+\bar{b}_D)$ & $\frac{4}{3}(3\bar{b}_0+\bar{b}_D)$ & $\frac{2}{3}(6\bar{b}_0+5\bar{b}_D+3\bar{b}_F)$ \\
    \hline\hline
  \end{tabular}

\end{table}

\subsection{Application to recent $n_f=2+1$ LQCD simulations}
At present, most LQCD simulations employ a single lattice spacing $a$ and take discretization effects as  systematic uncertainties.
A similar strategy has been adopted by theoretical studies. On the other hand, one may combine
 the LQCD simulations from different collaborations and perform a quantitative study of the discretization effects.
 Among the latest LQCD simulations,  several collaborations employed the $\mathcal{O}(a)$-improved or `clover' Wilson action, e.g. PACS-CS (with $a=0.0907(14)$ fm
 and $c_\mathrm{SW}=1.715$), QCDSF-UKQCD (with $a=0.0795(3)$ fm and $c_\mathrm{SW}=2.65$), HSC and NPLQCD (with $a_s=0.1227(8)$ fm,
 $a_t=0.03506(23)$ fm, $c^s_\mathrm{SW}=2.6$, and $c^t_\mathrm{SW}=1.8$) Collaborations. These simulations are performed at three different values of lattice spacing $a$ and with different light-quark masses and, therefore, in principle allow for a quantitative study of the discretization effects on the octet baryon masses.

It should be noted that both the HSC~\cite{Lin:2008pr} and the NPLQCD~\cite{Beane:2011pc} simulations  employed the anisotropic clover fermion action~\cite{Chen:2000ej}. In this action,  the temporal lattice spacing is chosen to be much smaller than the spatial lattice spacing. The EFT for such a LQCD setup has been worked out  in   Ref.~\cite{Bedaque:2007xg}, which in principle is more appropriate to be employed to study the HSC and NPLQCD simulations. On the other hand, this EFT has to introduce more LECs to discriminate the temporal and spatial lattice spacing effects. As we will see, present limited LQCD data do not allow us to perform such a study.  Therefore, in our study we assume that these simulations are performed with a single lattice spacing, $a_s$, and we treat the difference between $a_s$ and $a_t$ as higher-order effects.

As in Refs.~\cite{Ren:2013dzt,Ren:2013oaa}, we focus on the LQCD data from the above four collaborations with $M_{\pi}<500$ MeV and $M_{\phi}L>3.8$ to ensure the applicability  of  SU(3) covariant BChPT. In total, there are $12$ sets of LQCD data (each set includes the $N$, $\Lambda$, $\Sigma$, and $\Xi$ masses)  from the PACS-CS ($3$ sets), QCDSF-UKQCD ($2$ sets), HSC ($3$ sets), and NPLQCD ($4$ sets). In order to better ascertain the values of LECs, the experimental octet baryon masses are also included in the fits.

\begin{table}[t]
\small
\centering
\caption{Values of the LECs from the best fit to the LQCD data and the experimental data at
 $\mathcal{O}(p^4)$ with and without discretization effects.}
\label{Tb:fitcoef}
\begin{tabular}{lrr||lrr}
\hline\hline
                      &   BChPT   &  WBChPT  &  & BChPT  & WBChPT \\
\hline
  $m_0$~[MeV]         &  $910(20)$      &   $915(20)$    &  $d_1$~[GeV$^{-3}$]  &  $0.0295(124)$   &  $-0.0196(121)$  \\
  $b_0$~[GeV$^{-1}$]  &  $-0.579(56)$   &  $-0.557(50)$  &  $d_2$~[GeV$^{-3}$]  &  $0.342(65)$     &  $0.230(58)$     \\
  $b_D$~[GeV$^{-1}$]  &  $0.211(56)$    &  $0.201(48)$  &   $d_3$~[GeV$^{-3}$]  &  $-0.0314(63)$     &  $-0.0557(56)$ \\
  $b_F$~[GeV$^{-1}$]  &  $-0.434(43)$   &  $-0.359(41)$ &   $d_4$~[GeV$^{-3}$]  &  $0.372(114)$   &  $0.304(1008)$    \\
  $b_1$~[GeV$^{-1}$]  &  $0.730(10)$      &  $0.810(8)$  &  $d_5$~[GeV$^{-3}$]  &  $-0.401(110)$   &  $-0.237(88)$    \\
  $b_2$~[GeV$^{-1}$]  &  $-1.21(18)$     &  $-0.819(26)$  & $d_7$~[GeV$^{-3}$]  &  $-0.0913(58)$    &  $-0.104(48)$   \\
  $b_3$~[GeV$^{-1}$]  &  $-0.340(153)$    &  $-0.357(12)$ & $d_8$~[GeV$^{-3}$]  &  $-0.132(79)$    &  $-0.0417(67)$   \\
  $b_4$~[GeV$^{-1}$]  &  $-0.776(16)$     &  $-0.780(15)$  & $\bar{B}_1$~[GeV$^{-3}$]$\times 10^{-2}$& -- & $-0.121(103)$ \\
  $b_5$~[GeV$^{-2}$]  &  $-1.15(287)$   &  $-1.34(23)$   & $\bar{B}_2$~[GeV$^{-3}$]$\times 10^{-2}$& -- & $-0.467(109)$   \\
  $b_6$~[GeV$^{-2}$]  &  $0.778(390)$   &  $0.889(199)$  & $\bar{B}_3$~[GeV$^{-3}$]$\times 10^{-2}$& -- & $0.344(267)$   \\
  $b_7$~[GeV$^{-2}$]  &  $0.899(26)$      &  $0.787(14)$ & $\bar{X}$~~[GeV$^{-3}$]$\times 10^{-4}$& -- & $0.606(5723)$   \\
  $b_8$~[GeV$^{-2}$]  &  $0.627(37)$     &  $0.817(28)$     \\
  \hline
  $\chi^2$ & $30.0$  & $28.0$  & $\chi^2/{\rm d.o.f.}$ & $0.91$ & $0.97$ \\
\hline\hline
\end{tabular}
\end{table}

In the $\mathcal{O}(a)$-improved Wilson action  the Pauli term $a\mathcal{L}^{(5)}$ is eliminated. As a result,
discretization effects originate only from the $\mathcal{O}(am_q)$ and $\mathcal{O}(a^2)$ terms. Therefore,
 only the fourth-order tree-level diagrams contribute, while the leading order tree-level  diagram and the tadpole/one-loop diagrams
 do not contribute. In the end, the discretization effects,
\begin{eqnarray}
 m_B^{(a)} &=& m_B^{\mathcal{O}(am_q)} + m_B^{\mathcal{O}(a^2)}\nonumber\\
  &=& -8ac_{\rm SW}W_0\left(\xi_lM_{\pi}^2 + \xi_s(2M_K^2-M_{\pi}^2)\right) -16 a^2c_{\rm SW}^2W_0^2\bar{X},
\end{eqnarray}
only contain $4$ new independent combinations of LECs, i.e., $\bar{B}_1$, $\bar{B}_2$, $\bar{B}_3$, and $\bar{X}$. Together with the $19$ unknown LECs appearing in the octet baryon masses in the continuum, there are in total $23$ free LECs  that need to be fixed.\footnote{In our fits, we set $W_0$ at 1 GeV$^{3}$. Later a more proper value will be used to check the naturalness of the resulting LECs, $\bar{B}_1$, $\bar{B}_2$, $\bar{B}_3$, and $\bar{X}$.} As in Ref.~\cite{Ren:2012aj}, the meson decay constant is fixed at its chiral limit value $F_{\phi}=0.0871$ GeV.
For  the baryon axial coupling constants, we use $D=0.8$ and $F=0.46$~\cite{Cabibbo:2003cu}. The renormalization scale is set at $\mu=1$ GeV.

\begin{figure}[t!]
  \centering
  \includegraphics[width=\textwidth]{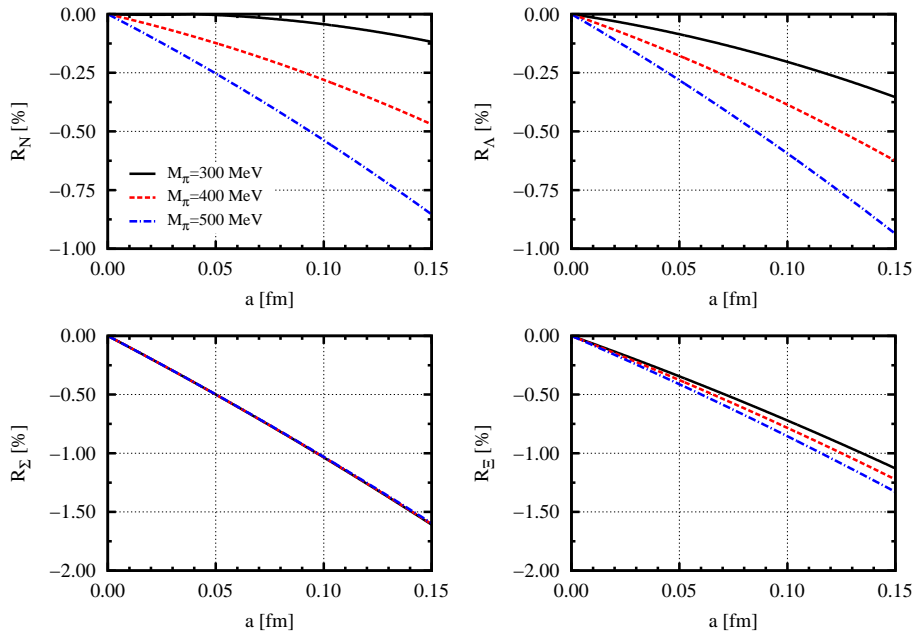}
  \caption{(color online). Finite lattice spacing effects on the octet baryon masses, $R_B=m_B^{(a)}/m_B$, as functions of lattice spacing $a$ for $M_{\pi}=0.3,~0.4$, and $0.5$ GeV, respectively. The SW coefficient is set at $c_{\rm SW}=1.715$, the value of the PACS-CS Collaboration. The strange quark mass is fixed at its physical value dictated by the leading order ChPT.}
  \label{Fig:adependence}
\end{figure}

In order to study the discretization effects on the octet baryon masses, we perform two fits. First, we use the continuum octet baryon mass formulas to fit the LQCD and experimental data. Second, the mass formulas of Eq.~(\ref{Eq:mass}) with discretization effects taken into account are employed to fit the same data. In both fits, the finite-volume corrections to the LQCD simulations are always taken into account self-consistently~\cite{Ren:2012aj}.  The LECs, together with the $\chi^2/\mathrm{d.o.f.}$, obtained from the two best fits are tabulated in Table~\ref{Tb:fitcoef}. It is clear that
the $19$ LECs  remain similar whether or not discretization effects are taken into account.
The total $\chi^2$ changes from $30$ for the first fit to $28$ for the second fit, indicating that the data can be described slightly better. On the other hand, the $\chi^2/\mathrm{d.o.f.}$ slightly increases from $0.91$ to $0.97$, implying  that discretization effects do not play an important role in describing the present LQCD data.\footnote{This is in contrast with the finite-volume effects. In Ref.~\cite{Ren:2012aj}, it is shown that a self-consistent treatment of finite-volume effects
is essential to obtain a $\chi^2/\mathrm{d.o.f.}$ about 1.}
 This justifies their treatments as systematic uncertainties without being taken into account explicitly in the fitting, as done in most previous theoretical and LQCD studies.
 It should be noted that the one-sigma uncertainties of the LECs $\bar{B}_1$, $\bar{B}_2$, $\bar{B}_3$, and, particularly, $\bar{X}$ are rather large. This shows clearly the need to perform LQCD simulations at multiple lattice spacings in order to pin down more precisely discretization effects, which has long been recognized~\cite{Kronfeld:2002pi}.
\begin{table}[h!]
  \centering
  \caption{Extrapolated octet baryon masses (in units of MeV) to the physical point with the LECs determined by fitting to the LQCD data alone.}
  \label{Tab:extrapolation}
  \begin{tabular}{cccc}
    \hline\hline
      & BChPT & WBChPT & Exp.~\cite{Beringer:1900zz} \\
    \hline
    $\chi^2/\mathrm{d.o.f.}$ & $0.89$  & $1.0$ & -- \\
    \hline
    $m_N$ & $889(21)$ & $865(39)$ & $940(2)$ \\
    $m_\Lambda$ & $1113(17)$ & $1087(41)$ & $1116(1)$ \\
    $m_\Sigma$ & $1163(19)$ & $1139(42)$ & $1193(5)$ \\
    $m_\Xi$ & $1333(16)$ & $1309(41)$ & $1318(4)$ \\
    \hline
    \hline
  \end{tabular}

\end{table}

In the above fits we have included the experimental data to better constrain the large number of LECs appearing at N$^3$LO. We can of course drop the experimental data, redo the fit, and calculate the octet baryon masses at the physical point. Such a procedure should be taken with caution, however, for the following reasons. First, we have a large number of unknown LECs (about 20). Second, the lightest LQCD data point has a $M_\pi$ about $300$ MeV, and it is still a bit away from the physical point. Third, all the $\chi^2/\mathrm{d.o.f.}$ are close to 1. These factors can make the extrapolations unstable with respect to moderate changes of the LECs. In Table \ref{Tab:extrapolation}, we tabulate the extrapolated octet baryon masses with two sets of LECs, determined from the fits in which finite lattice spacing effects are either taken into account or neglected. It is clear that the extrapolated masses agree within uncertainties, and so do the corresponding LECs (not shown). Nevertheless, the extrapolated nucleon mass still deviates about 60-80 MeV from its physical value, calling for LQCD simulations with smaller light-quark masses (than studied in the present work).

In Fig.~\ref{Fig:adependence}, we show the evolution of discretization effects as a function of the lattice spacing for three different pion masses with the  relevant LECs determined from the second fit.  It is seen that the discretization effects increase almost linearly with increasing lattice spacing $a$ for fixed pion mass. For fixed $a$, they increase with increasing pion mass as well. Furthermore, essentially no curvature is observed. It is clear that in our present work  the $\mathcal{O}( a m_q)$ terms dominate over the $\mathcal{O}(a^2)$ terms.  It should be stressed that the LEC $\bar{X}$ is consistent with zero and a fit without the $\mathcal{O}(a^2)$ contributions would have yielded very similar results as shown in Table ~\ref{Tb:fitcoef} and Fig.~\ref{Fig:adependence}.
For a lattice spacing up to $a=0.15$ fm, the finite lattice spacing effects on the baryon masses are less than 2\%, consistent with the LQCD study of Ref.~\cite{Durr:2008rw}.

The above results can be naively understood in the following way. Recall that 
$\frac{m_q}{\Lambda_\mathrm{QCD}}\sim a \Lambda_\mathrm{QCD}$ in our power-counting scheme. If we take $m_s=100$ MeV, $\Lambda_\mathrm{QCD}=300$ MeV, and $a=0.1$ fm, we obtain
$\frac{m_q}{\Lambda_\mathrm{QCD}}\approx0.3$ and $a{\Lambda_\mathrm{QCD}}\approx 0.15$. If we further assume that all the LECs are of natural size, i.e., $\sim1$, we then expect $\mathcal{O}(m_q^2):\mathcal{O}(am_q):\mathcal{O}(a^2)=4:2:1$. Remember that the quark masses are larger than their physical values while the lattice spacing is fixed to be around $0.1$ fm in the LQCD simulations, our actual numerical results seem to support this naive argument. Furthermore, we would like to point out that the $a$-dependent LECs $\bar{B}_1$, $\bar{B}_2$, $\bar{B}_3$, and $\bar{X}$ are of natural size. The values in Table \ref{Tb:fitcoef} appear to be small because we have set the dimensional quantity $W_0$ to be 1 $\mathrm{GeV}^3$. Its  more `proper' value can be estimated by noting the following relations $W_0 a\sim B_0 m_q$ and  $M_\pi^2\propto 2 B_0 m_q$ (in the leading-order ChPT), which yields $W_0\approx 0.02$ GeV$^3$. With this value, the LECs turn out to be $\bar{B}_1=-0.0605$ GeV$^{-3}$, $\bar{B}_2=-0.234$ GeV$^{-3}$, $\bar{B}_3=0.172$ GeV$^{-3}$, and $\bar{X}=0.152$ GeV$^{-3}$, which are of natural size as expected.

\section{Conclusions}\label{SecIV}

We have studied discretization effects on the octet baryon masses.
The $a$-dependent chiral Lagrangians are formulated for the first time in the SU(3) one-baryon sector and
discretization effects on the octet baryon masses are calculated for the unmixed Wilson action  up to $\mathcal{O}(a^2)$. By taking into account discretization effects and finite-volume corrections, we have performed a simultaneous fit of all the $n_f=2+1$ LQCD simulations, which are performed using the $\mathcal{O}(a)$-improved Wilson fermion action. We found that taking into account discretization effects can slightly improve the description of the LQCD octet baryon masses, but their effects are small. Furthermore, the values of the $19$ LECs appearing in continuum ChPT up to $\mathcal{O}(p^4)$ do not change much. Our studies showed that the treatment of discretization effects  as systematic uncertainties in the previous studies of the LQCD octet baryon masses seems to be justified.

With the LECs of Wilson ChPT fixed from the best fit, we have also studied the evolution of discretization effects with the lattice spacing and the pion mass.
It was shown that the discretization effects on the octet baryon masses are less than $2\%$ for lattice spacings up to $0.15$ fm, in
agreement with other LQCD studies.

Nevertheless, future lattice simulations performed at multiple lattice spacings will be extremely valuable to pin down
more precisely discretization effects (on the octet baryon masses) and to check the validity of  Wilson ChPT in the one-baryon sector.

\begin{acknowledgements}
X.-L.R thanks Dr. Hua-Xing Chen for useful discussions and acknowledges support from the Innovation Foundation of Beihang University for Ph.D. Graduates. L.S.G acknowledges support from
the Alexander von Humboldt foundation. This work was partly supported by the National Natural
Science Foundation of China under Grants No. 11005007, No. 11035007, and No. 11175002, the
New Century Excellent Talents in University Program of Ministry of Education of China under
Grant No. NCET-10-0029, the Fundamental Research Funds for the Central Universities, the
Research Fund for the Doctoral Program of Higher Education under Grant No. 20110001110087.
\end{acknowledgements}

\bibliographystyle{apsrev4-1}
\bibliography{EOMS-LA}

\end{document}